# Was there a COVID-19 harvesting effect in Northern Italy?

Augusto Cerqua[a], Roberta Di Stefano[b], Marco Letta[a], Sara Miccoli[b]

[a] Department of Social Sciences and Economics, Sapienza University of Rome
[b] Department of Methods and Models for Economics, Territory and Finance, Sapienza University of Rome

## Abstract

*We investigate the possibility of a harvesting effect, i.e. a temporary forward shift in mortality, associated with the COVID-19 pandemic by looking at the excess mortality trends of an area that registered one of the highest death tolls in the world during the first wave, Northern Italy. We do not find any evidence of a sizable COVID-19 harvesting effect, neither in the summer months after the slowdown of the first wave nor at the beginning of the second wave. According to our estimates, only a minor share of the total excess deaths detected in Northern Italian municipalities over the entire period under scrutiny (February - November 2020) can be attributed to an anticipatory role of COVID-19. A slightly higher share is detected for the most severely affected areas (the provinces of Bergamo and Brescia, in particular), but even in these territories, the harvesting effect can only account for less than 20% of excess deaths. Furthermore, the lower mortality rates observed in these areas at the beginning of the second wave may be due to several factors other than a harvesting effect, including behavioral change and some degree of temporary herd immunity. The very limited presence of short-run mortality displacement restates the case for containment policies aimed at minimizing the health impacts of the pandemic.*

**JEL-Codes:** C21; C53; J11

# 1. Introduction

At the onset of the COVID-19 pandemic, there was much speculation in Italy as elsewhere about a potential 'harvesting effect' of COVID-19, i.e. a short-term increase of mortality later followed by a corresponding decrease in deaths. The claim was that COVID-19 fatalities, whose median age was around 80 years, were, for the vast majority, very vulnerable people who, in the absence of the pandemic, would have 'died anyway' shortly after they actually did, due to other causes. According to the proponents of this hypothesis, COVID-19 excess mortality would have been largely re-absorbed in the months after the mortality peak, as the virus would have simply anticipated a large number of occurred deaths. Stated differently, this implies that when the spread of COVID-19 progressively slows down, one should observe a significant reduction in mortality, which would counterbalance the abnormal increases experienced during the peak.

Albeit this position has been quickly picked up by COVID-19 'skeptics' to protest against the social distancing policies introduced by most governments,[1] research on the plausibility of this claim is still scarce as documented in Section 2, both due to lags in data availability and the short time-span observable so far. We investigate the possible harvesting effects of the COVID-19 pandemic by looking at the excess all-cause mortality trends in Northern Italy, one of the areas with the highest COVID-19 death toll in the world. Specifically, we employ the data-driven methodology introduced by Cerqua et al. (2020) to estimate excess mortality and then investigate how it has evolved during three separate periods: i) the peak of the first wave in Italy, February - May 2020; ii) the 'summer break', i.e. the tail of the first wave, going from June to September; and iii) the beginning of the second wave, i.e. October and November 2020.

During the summer break, a 'negative' excess mortality is detected. Nevertheless, this reduction in *observed* mortality compared to the counterfactual mortality figures *predicted* by our model is far too limited (corresponding to 16% of the total excess deaths observed during the first wave) to compensate for the abnormally high excess mortality of the first wave. During the onset of the second wave, new excess mortality clusters are detected, and we estimate that, as a consequence, the harvesting effect further shrinks to less than 12% when considering the entire February - November period. Still, we do observe a negative and statistically significant spatial autocorrelation between the mortality patterns of the two waves in some areas (the provinces of Bergamo and Brescia, in particular) where, remarkably, we also detect a 'negative' excess mortality not only during the summer months but also in October and November. This is consistent with the well-documented lower incidence of COVID-19 during the beginning of the second wave in these Lombardy provinces, which even led some mayors to ask to be exempted from the November 3, 2020 decree, which imposed a 'red zone' in the entire region. While such reversal of patterns between the two waves could be due to several factors, such as a behavioral change and some degree of herd immunity in the most affected areas, it is by no means enough to compensate for the mortality boom of the first wave. Indeed, even in these hardest-hit areas, the harvesting effect can only explain up to 17% of the COVID-19 related deaths experienced during the entire period under scrutiny.

---

[1] In Italy, early in the pandemic, there was a fierce debate, which featured heavily in media outlets, about whether COVID-19 victims had died 'with coronavirus' or 'from coronavirus'.

On top of these period-by-period comparisons, we also compute the cumulative number of deaths over the entire period under scrutiny (February - November 2020), which sums to 49,816 deaths more than 'expected' in Northern Italy. This corresponds to an increase in mortality of +20% with respect to an 'ordinary' year, i.e. in a 'no-COVID' counterfactual scenario. Overall, this evidence suggests that, although COVID-19 has probably anticipated the death of some of the frailest individuals of the Italian population, in the vast majority of cases, it killed relatively healthy people who did not have a short life expectancy before the pandemic's arrival.

## 2. The harvesting effect

The harvesting effect, or mortality displacement, is identified as an increase in deaths followed by fewer deaths than expected after the mortality crisis. During exogenous shocks such as heat waves or cold spells, the selective mortality among the frailest individuals increases the deaths among the total population and leaves a relevant proportion of strong survivors (Luy et al. 2020). After the shock, the number of deaths is below the expected number, and, therefore, a compensation in mortality can be observed between the crisis and the following period (Toulemon and Barbieri 2008).

Several scholars studied the harvesting effect caused by particular events, such as heat waves or cold spells (e.g. Baccini, Kosatsky and Biggeri 2013; Cheng et al. 2018; Grize et al. 2005; Qiao et al. 2015; Stafoggia et al. 2009; Toulemon and Barbieri 2008), seasonal influenza (e.g. Lytras et al. 2019) or air pollution (Rabl 2005). Lytras et al. (2019) found out that the influenza A(H1N1)pdm09 affected the frailties individuals that would have died in the short-term because of other causes, while influenza A(H3N2) and type B caused an excess of influenza deaths among people who would not have died in the same year. Stafoggia et al. (2009), studying deaths that occurred in Rome between 1987-2005, figured out that high levels of mortality during winter periods can reduce the effect of heat waves on mortality compared with years of winters with low levels of mortality.

To our knowledge, only a few papers assess the potential presence of the harvesting effect during the COVID-19 pandemic. Rivera et al. (2020) stated that in the US, the very high mortality due to COVID-19 spans over a more extended period than other influenza or pandemic, and probably no harvesting would be observed in periods following the worst waves of the COVID-19 pandemic. In fact, in a study that analyzes the 2020 life expectancy decrease in the US, Andrasfay and Goldman (2021) did not find evidence of a harvesting effect due to COVID-19. Alicandro et al. (2020) indicated the possible presence of a harvesting effect at the end of the first wave of the pandemic in Italy, except for the Lombardy region, where this effect was less pronounced at that time. Similarly, Scortichini et al. (2020), by analyzing excess mortality across Italian provinces, suggested the possible presence of harvesting effect in some areas of Central and Southern Italy at the end of the first wave of the pandemic. The Italian National Institute of Statistics (Istat) and Istituto Superiore di Sanità (ISS) (2020b) reported some evidence of harvesting effect in some areas of Northern Italy during the summer months when the infections were minor. In contrast, the recent study by Canoui-Poitrine et al. (2021), who estimate the number of excess deaths among nursing home residents during the first wave of the pandemic in France, finds no evidence harvesting effects up to the end of August.

### 3. Data and methodological approach

To determine the potential presence of a harvesting effect in Northern Italy, we first estimate the excess all-cause mortality due (directly or indirectly) to the COVID-19 pandemic at the municipality-level and then investigate its evolution over time across the three different periods described above. The first step is made necessary by the lack of reliable data on the deaths caused by COVID-19, especially at a disaggregated level. Indeed, official data on the death toll of COVID-19 at the local level are scarce,[2] and they are likely to suffer from substantial underreporting (Ghislandi et al., 2020).

Excess mortality is defined as the difference between the observed mortality in the presence of a pandemic and the counterfactual scenario of mortality in the pandemic's absence. It includes the number of deaths due directly to COVID-19 infections as well as the deaths due indirectly to COVID-19, i.e. the collateral effects of the lockdown. During the lockdown, the likelihood of dying for road[3] and workplace accidents, pollution-related diseases, or criminal activities decreases. At the same time, the likelihood of dying for the stress on the public health system increases. The estimation of excess mortality is made possible thanks to the data released on February 3, 2021, by Istat on the number of daily certified deaths for the period January 1, 2015 – November 30, 2020, for all Italian municipalities.[4]

---

[2] In Italy, official data on SARS-CoV-2 reported cases are released only at the provincial level (the number of infected people) or at the regional level (the number of COVID-19 deaths).

[3] The Istat - ACI (2020) report records a decrease in victims due to road accidents in the period January - September 2020 of 1,788 (-26.3%). The percentage increases to 34% by considering the period January - June 2020.

An accurate estimation of excess mortality requires the construction of a reliable counterfactual scenario. In the context of the pandemic mortality estimation, different approaches were used.[5] The most common is what we call the 'intuitive' approach. It consists of using the simple average of the numbers of deaths observed for the same unit in the past. This approach has been adopted by several national and international institutions and employed in many scientific works. It is a simple approach that does not employ any model, but it may provide excess mortality estimates which are too sensitive to outliers. Another possible approach is the use of the counterfactual approaches, such as the difference-in-differences or the synthetic control method estimators. However, these approaches are ill-suited in a setting where it is hard to find plausible control groups, i.e. municipalities potentially not affected (directly or indirectly via containment measures) by the pandemic for several months.

An attractive methodological solution to such an estimation problem is the recently developed machine learning control method (MLCM) inspired by the train-test-treat-compare process proposed by Varian (2016). In the context of COVID-19, the MLCM can be applied by drawing on the predictive ability of ML algorithms to generate a no-COVID counterfactual scenario for each unit by using exclusively pre-pandemic information (Cerqua and Letta, 2020). In our setting, the use of the MLCM is made possible by constructing a comprehensive time-series cross-sectional database on Italian municipalities.

The reason to prefer MLCM over the 'intuitive' approach lies in its ability to estimate more accurate counterfactual scenarios. Cerqua et al. (2020) demonstrate that considering the Mean Squared Error (MSE), on average, there is a sizable gain in terms of estimation accuracy compared with the intuitive estimates, especially for small and medium-sized municipalities. For this reason, we investigate the presence of the harvesting effect on Northern Italy by applying the MLCM approach used by Cerqua et al. (2020) to retrieve excess mortality estimates at the municipality-level. The mortality scenario without the pandemic, i.e. the cumulative number of deaths per 10,000 inhabitants in an ordinary situation, is estimated using 16 selected covariates from 2015 to 2019, including the demographic, health system, economic, and contamination (air pollution) features.[6]

---

[4] Due to the creation of Mappano as a new administrative unit in 2017 and to the lack of mortality data for all years, we cannot analyze six municipalities: Borgaro Torinese, Caselle Torinese, Leini, Mappano, and Settimo Torinese. Besides, as 2020 is a leap year, we decided to ignore the deaths on February 29 for comparability with data from previous years.

[5] See Section 2 of Cerqua et al. (2020) for a review of the methodologies used to estimate excess mortality during pandemics.

For each considered period (in our case, the peak of the first wave, the summer break, and the beginning of the second wave), we train and test our random forest algorithm on the pooled 2015-2019 (on which, as typical in the ML literature, we apply a random split and use 80% of the full sample as the training set and the remaining 20% as the testing sample) dataset to predict, for the 2020 sample, estimates of local mortality in a counterfactual scenario without the pandemic. It is then easy to retrieve excess mortality as the difference between observed and predicted mortality.

Cerqua et al. (2020) use three ML algorithms: Least Absolute Shrinkage and Selection Operator (LASSO), random forest, and stochastic gradient boosting. In this work, we apply the ML using the random forest algorithm, a fully non-linear technique based on the aggregation of many decision trees (1000, in our case), as Cerqua et al. (2020) demonstrate that it performs well for all municipality sizes. The choice to circumscribe the analysis only on Northern Italy is dictated by the fact that it was the epicenter of the pandemic in Italy during the peak of the first wave (February - May 2020) and one of the mortality hotspots in Europe. As such, we deem it a representative case study to test for a potential COVID-19 harvesting effect.[7] We investigate how all-cause deaths have evolved during three separate periods: i) the peak of the first wave in Italy, February - May 2020; ii) the 'summer break', i.e. the tail of the first wave, going from June to September; iii) the beginning of the second wave, October and November 2020, according to the division made by the fourth report Istat, ISS (2020a).

We do so by showing the choropleth maps of each separate period as well as by using one of the most important indexes for studying spatial relationships: the Moran's I. Moran's I can be of two types: the global bivariate Moran's I and local bivariate Moran's I (bivariate Local Indicators of Spatial Association, or more simply bivariate LISA). The former provides summary statistics for overall spatial clustering. It varies between +1 and -1: a value close to +1 indicates a strong positive spatial autocorrelation. Otherwise, a value close to -1 reveals that the spatial autocorrelation is negative, while 0 indicates a random spatial pattern. The bivariate LISA is instead applied to depict the spatiality of how the value of one variable is surrounded by values of a second variable (Anselin 1995). Basically, the bivariate LISA measures the relationships between spatial units and their neighboring spatial units and maps statistically significant clusters of the phenomena under analysis. The neighboring structure across municipalities is measured by a spatial weights matrix based on the inverse geographical (Euclidean) distances between municipalities' centroids.[8] The weight matrix is then standardized such that its rows sum to unity (in order to compute neighborhood averages) and have zeros along the leading diagonal (see Maddison, 2006).

---

[6] The full list of covariates is the following: the share of men in the population, the share of those aged 65+ (overall as well as only men), the share of those aged 80+ (overall as well as only men), the resident population, the overall number of deaths in the previous year, the overall number of deaths in the 7 weeks before the COVID-19 outbreak in Italy, the number of employees, the share of employment in manufacturing, the PM-10 as a measure of air quality, the population density, the degree of urbanization of the municipality, the dummy of the presence of a hospital in the municipality, the dummy of the presence of a hospital in at least one of the neighboring municipalities, and the number of deaths due to road accidents in the previous year. For more details, see Cerqua et al. (2020).

[7] On the contrary, the excess mortality observed during the peak of the first wave in Central and Southern Italy could be too mild and uneven to determine a harvesting effect.

Thanks to the bivariate LISA, we will identify the following types of association: positive autocorrelation, which occurs where high values of variable 1 are surrounded by high values of variable 2 (high–high hotspots, HH) or where there is a concentration of low values (low–low coldspots, LL); or negative spatial autocorrelation, namely places where low values of variable 2 surround high values of variable 1 (High–Low clusters, HL), or vice versa (Low-High clusters, LH). As in Frigerio et al. (2015), we will use 999 random permutations to determine the statistical significance for each cluster.

In our analysis, we will use the global bivariate Moran's I to study the overall spatial correlation of excess mortality values of the first wave on the summer break (second wave) in Northern Italy and the bivariate LISA to measure the clustering patterns of excess mortality values of the first wave and the summer break (second wave). We will investigate whether the patterns of similarity and dissimilarity in the clustering of excess mortality values remained stable across the three time periods.

## 4. Results

### 4.1 Excess mortality estimates

The excess mortality estimates from all-cause deaths relative to the first phase of the pandemic, the so-called peak of the first wave from February 21 to May 31, are shown in Figure 1 for Northern Italy. Compared with the counterfactual scenario, the municipalities with the highest excess mortality are located in the provinces of Bergamo, Brescia, Cremona, Lodi in the Lombardy region. Quite impressively, 40.9% of the Lombardy municipalities recorded excess mortality of over 100%. Wide clusters of municipalities with excess mortality above 100% are also present in Piacenza and Parma provinces in the Emilia Romagna region and the Lombardy region. Clusters of municipalities with an excess of deaths over 50% are located in Milan, Mantova and Pavia (Lombardy), again in Piacenza and Parma provinces (Emilia Romagna), but also in the provinces of Imperia (Liguria), Cuneo and Alessandria (Piedmont), and Trento (Trentino Alto-Adige). In many municipalities of the Liguria region and the provinces of Turin (Piedmont), Reggio Emilia, Rimini, and Forlì-Cesena (Emilia Romagna), the excess mortality is between 20% and 50%. During the first wave, 126,896 deaths were recorded in Northern Italy, and we estimate 41,586 excess deaths in this period. This corresponds to an increase of +48.7% in the number of deaths due directly or indirectly to the pandemic during the peak of the first wave in Northern Italy.

---

[8] The distance threshold is 15.1 km, which is the minimum threshold in order to avoid neighborless municipalities.

In the summer break from June 1 to September 30, defined by Istat, ISS (2020a) as a transition phase, there are few municipalities with excess mortality over 50%, as displayed in Figure 2. Most of the municipalities in Northern Italy do not record excess mortality, and small clusters of municipalities with an excess of deaths above 50% are located in the Aosta Valley region and Turin and Cuneo provinces in Piedmont. Our estimates confirm the evolution described by Istat, ISS (2020a), which has connected the presence of 'negative' excess mortality with the lower number of deaths recorded in this period in comparison with the average deaths in the years 2015-2019. During the summer break, 94,382 deaths were recorded in Northern Italy, 6,608 less than predicted by our random forest model for a no-COVID counterfactual scenario. This evidence might suggest a moderate presence of harvesting effect during the summer months. However, this reduction in the number of all-cause deaths is extremely limited with respect to the massive number of deaths observed in the first wave, as it only accounts for less than 16% (6,608/41,586) of the total excess deaths of the first wave.

At the onset of the second wave, many Northern Italy municipalities register a relevant excess of deaths, but the geographical pattern is different from what was observed during the first wave of the pandemic. As shown in Figure 3, broad clusters of municipalities with excess mortality above 100% are located in Cuneo (Piedmont) and Belluno (Veneto), and the Aosta Valley region. In various provinces of Piedmont, Trentino Alto-Adige, and Lombardy, many municipalities have an excess of deaths over 50%. Notably, the areas of Lombardy which had been most affected during the first wave of the pandemic, such as the provinces of Bergamo, Brescia, Cremona, and Lodi, experience low or even no levels of excess mortality. In the province of Milan, which had already been severely affected by the virus during the first wave, there are many municipalities with excess mortality of over 20%. In the most affected areas of Emilia Romagna during the first wave, namely Piacenza and Parma provinces, most municipalities record negative excess mortality or low levels, while in areas surrounding Bologna, more municipalities record an excess of deaths over 20% in comparison to the first wave. In Liguria, a cluster of municipalities with a mortality excess above 50% is located around Genova. In October and November 2020, there were 67,865 deaths in Northern Italy, and we estimate 14,838 excess deaths, i.e. an increase of +28.0% with respect to a 'no-COVID' scenario. For Northern

Italy as a whole, therefore, the harvesting effect during the entire February - November period can only account for a small portion of the total excess deaths detected.

At the beginning of the second wave, the excess mortality is lower in some particular areas harshly hit by the deaths' increase during the first wave, and it is exceptionally high in areas that did not experience a very high excess mortality in the first period of the pandemic. During the first wave, the areas with the highest rate of infection were well-defined and less widespread in comparison to the summer break and the second wave.

**Figure 1** - Percentage of municipal excess deaths detected during the peak of the first wave with respect to the counterfactual scenario estimated via random forest

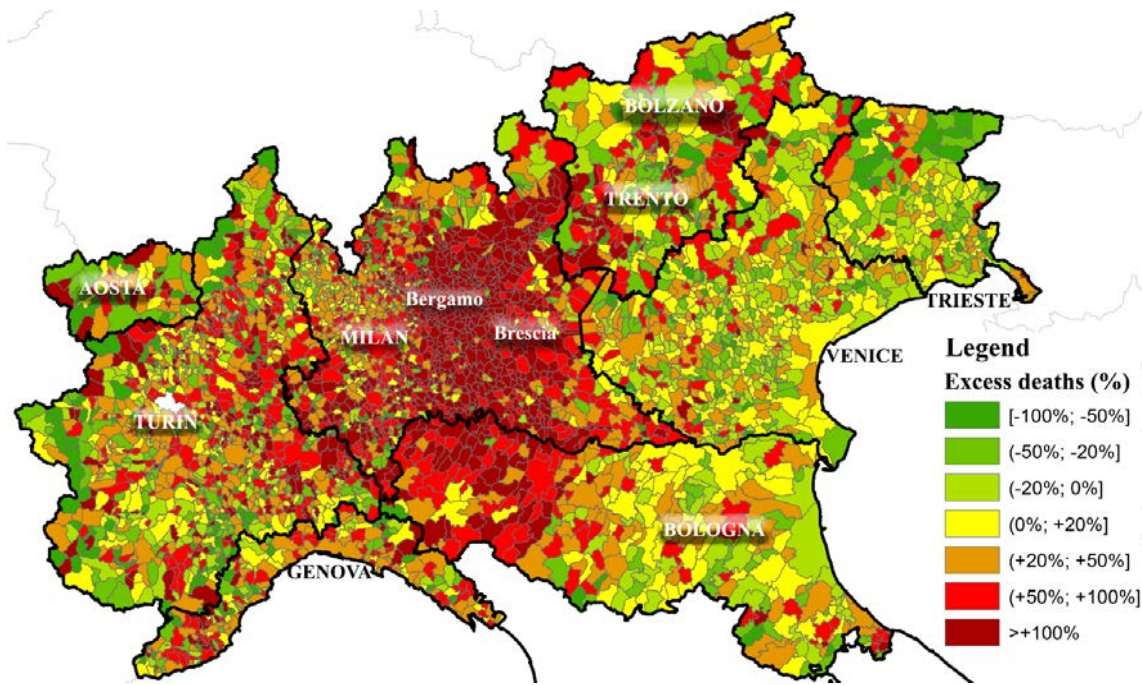

**Figure 2** - Percentage of municipal excess deaths detected during the 'summer break' with respect to the counterfactual scenario estimated via random forest

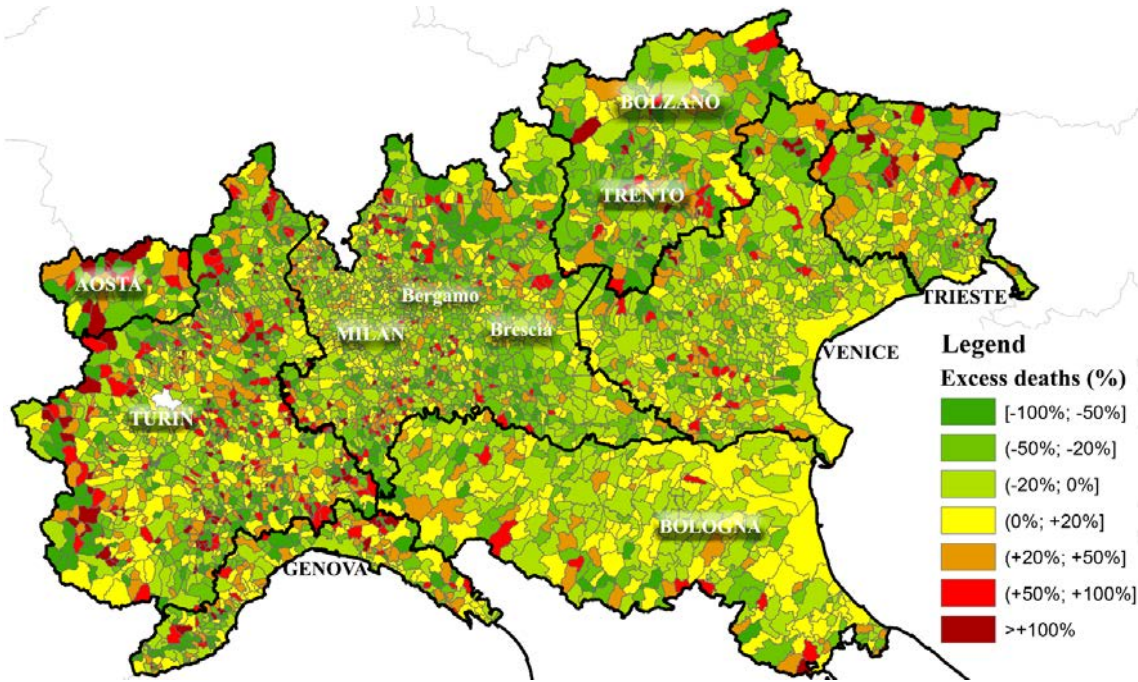

**Figure 3** - Percentage of municipal excess deaths detected during the onset of the second wave with respect to the counterfactual scenario estimated via random forest

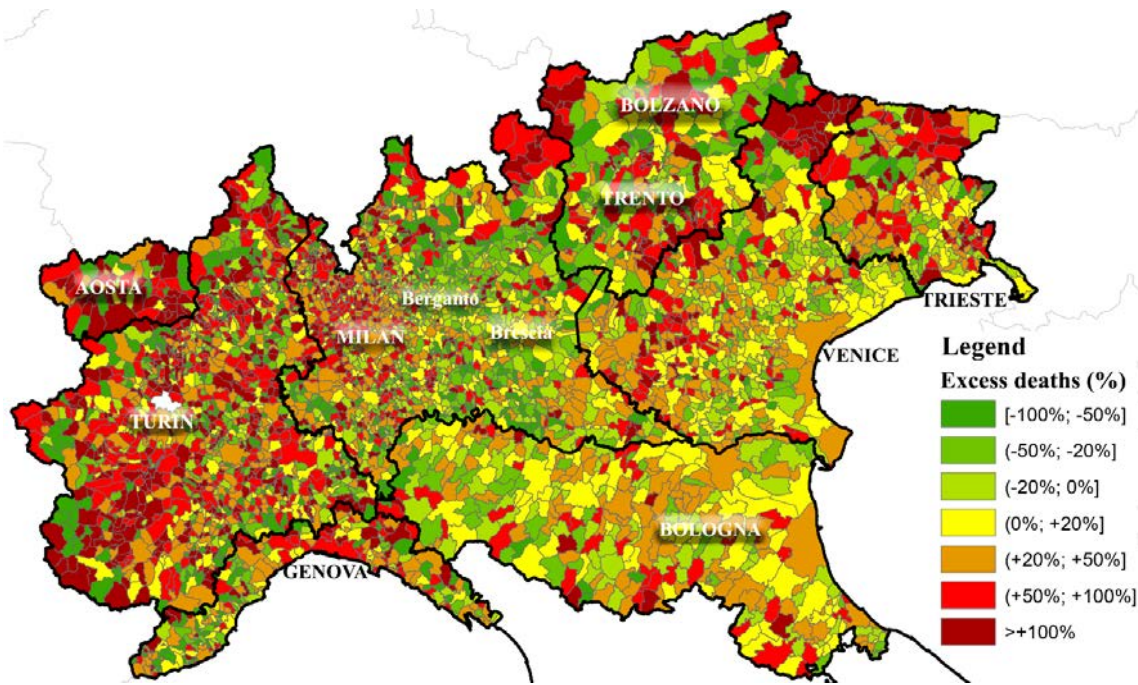

### 4.2 Spatial correlation indexes

Our starting point is the spatial correlation between the excess mortality values of the first wave and the summer break. The global bivariate Moran's I statistic is -0.053, indicating low negative spatial autocorrelation. This means that, on average, in the North of Italy, the spatial association between excess mortality values of the first wave and the summer break has only a small degree of spatial clustering. However, the global statistic can mask substantial local variation in spatial autocorrelation. Hence, we also computed bivariate LISA to map spatial autocorrelation for each municipality. The resulting maps are displayed in Panel A of Figure 4, which maps bivariate LISA clusters, and in Panel B of Figure 4, which shows their statistical significance at the 5%, 1%, or 0.1% significance levels. Overall, this analysis demonstrates that there are a few distinct geographic patterns of spatial clustering.

In the map in Panel A, the HH hotspots (dark red) are areas where municipalities with higher-than-average excess mortality in the first wave are surrounded by municipalities with higher-than-average excess mortality in the summer break. As evident from the map, very few municipalities exhibit high excess mortality in both considered periods.

The high–low (HL) clusters (salmon-colored) are areas where municipalities with high excess mortality in the first wave have neighboring municipalities with low excess mortality in the summer break. These clusters are most prominent in the Bergamo and Brescia provinces, harshly hit by the virus during the first phase of the pandemic.

The low–high (LH) and low–low (LL) clusters also demarcate places of bivariate extremes. The LH clusters represent areas with low excess mortality in the first wave, with neighboring municipalities with high excess mortality in the summer break. LH clusters are present in Aosta Valley and Piedmont, while tracts in LL clusters appear primarily in the areas around Trento. These places have the lowest levels of excess mortality in the first wave as well as in the summer break.

We then investigate the spatial correlation between the excess mortality values of the first wave and the onset of the second wave. The global bivariate Moran's I statistic stays negative and low (-0.091). We then use the bivariate LISA to identify clusters of the excess mortality values of the first and second waves and report them in Figure 5. Two relevant patterns emerge: i) some areas which were only moderately hit during the first wave experienced high levels of excess mortality in October and November. In particular, these areas are concentrated in the provinces of Varese, Como, and Milan in Lombardy, Belluno in Veneto, Udine in Friuli-Venezia Giulia, Cuneo and Biella in Piedmont; ii) the municipalities surrounding Bergamo and Brescia, the most harshly hit during the first wave in Italy, exhibit low levels of excess deaths at the beginning of the second wave.

**Figure 4** - Bivariate LISA of excess mortality values of the peak of the first wave and the summer break. **Panel a** Cluster map and **Panel b** Cluster significance

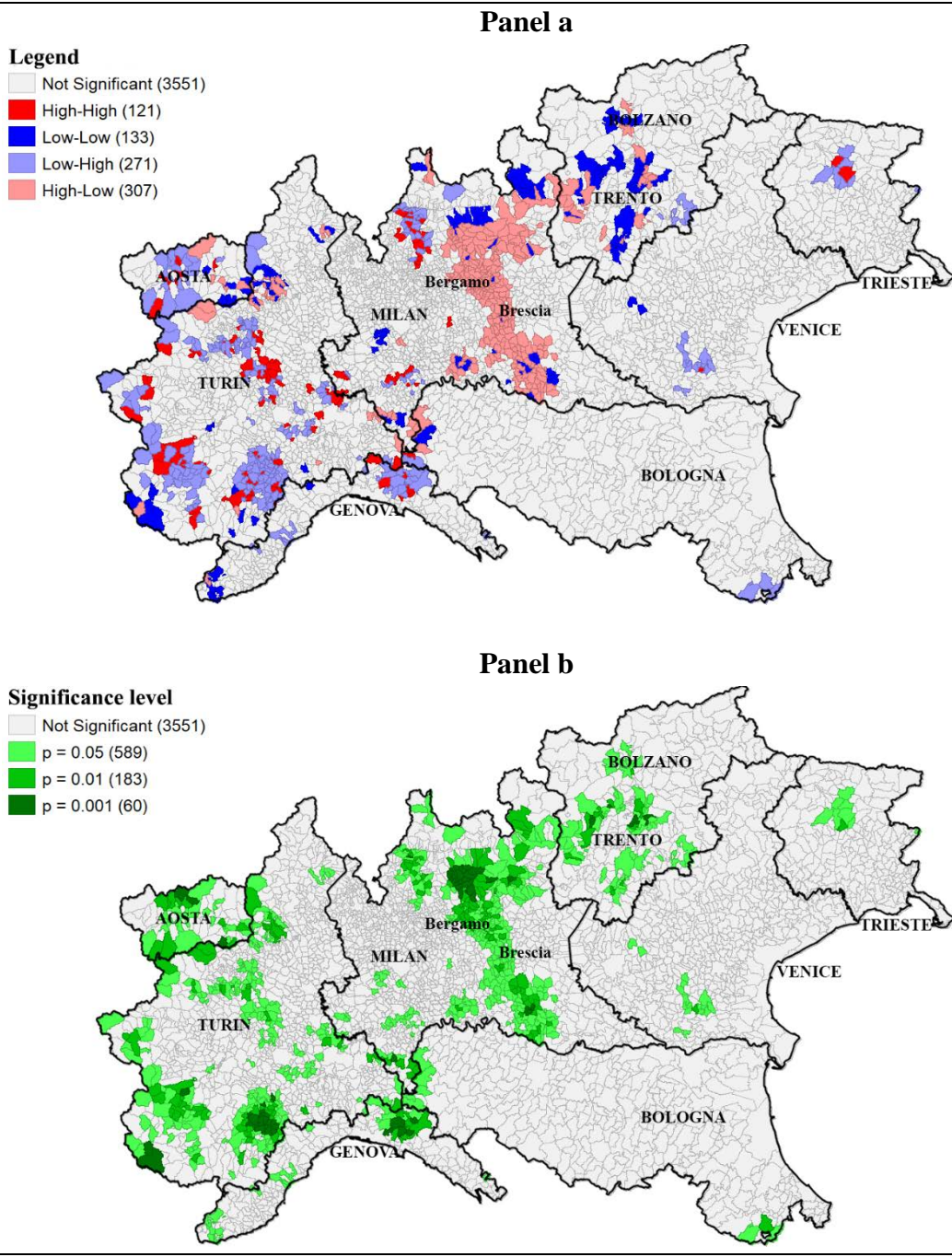

**Figure 5** - Bivariate LISA of excess mortality values of the peak of first wave and the onset of the second wave. **Panel a** Cluster map and **Panel b** Cluster significance

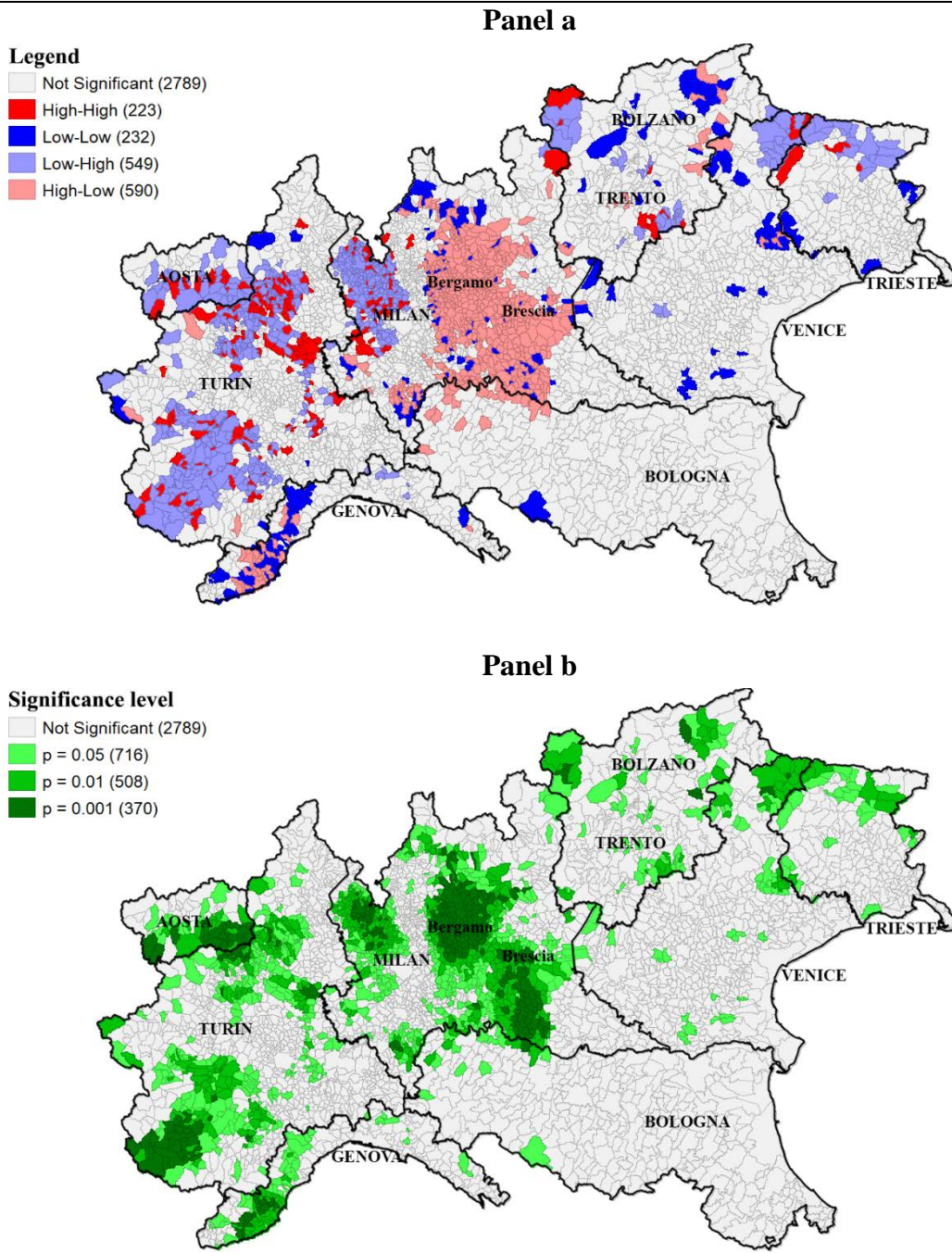

Overall, while our examination of overall trends in excess deaths for Northern Italy suggests very limited evidence of relevant harvesting effects, our spatial analysis gives compelling evidence that the areas of Northern Italy which were -hit the hardest in the first phase of COVID-19, then experienced a decrease in the number of deaths of a larger magnitude, and over a longer time-span, with respect to the majority of the other Northern Italy municipalities. Let us, therefore, take a closer look at excess mortality dynamics in these most affected areas. By focusing on the provinces of Bergamo and Brescia, in the first wave, we observed excess mortality of +164% (16,754 individuals died in front of an 'expected' number of deaths of 6,351), during the summer break, a drop in the number of deaths of -17.2% (6,579 individuals died in front of an 'expected' number of deaths of 7,948) and at the onset of the second wave a drop in the number of deaths of -5.1% (3,936 individuals died in front of an 'expected' number of deaths of 4,147). Remarkably, in these areas, the all-cause mortality balance sign is negative even at the beginning of the second wave. This evidence suggests a somewhat more pronounced harvesting effect in the most affected areas of Northern Italy during the first wave. Of the 10,403 excess deaths that occurred during the first wave in these areas, we estimate that 1,580 individuals would have died anyway by the end of November 2020. However, this still means that over 83% of the deaths due (directly or indirectly) to COVID-19 concern relatively healthy people that did not have a short life expectancy before the pandemic's arrival.

## 5. Conclusion

By studying mortality dynamics in the immediate aftermath of the first COVID-19 wave in Northern Italy, one of the hardest-hit territories of the world, we find only limited evidence of a COVID-19 harvesting effect. The impressive COVID-19 first-wave excess mortality in Northern Italian municipalities was only marginally 'compensated for' by a subsequent decline in mortality. In line with Canoui-Poitrine et al. (2021) findings for nursing home residents in France, we do not find that COVID-19 only affected those whose health was already inevitably compromised. The vast majority of COVID-19 deaths are not 'anticipated' deaths but sudden and 'unexpected' ones.

We document a slight reduction in total mortality during the summer months and new excess mortality clusters at the beginning of the second wave. When considering these dynamics jointly, for Northern Italy as a whole, the harvesting effect can account only for a minor share of the total excess deaths detected over the entire period. We also detect a statistically significant and negative spatial autocorrelation between the mortality trend of the first wave and that of the second, and a negative mortality balance at the beginning of the second wave, in some territories such as the provinces of Bergamo and Brescia. In these areas, the most severely affected ones during the first wave, less than 20% of the COVID-related deaths might have occurred anyway by the end of November 2020. However, these inverse dynamics are likely the joint outcome of a combination of causal factors, such as some degree of temporary herd immunity coupled with long-lasting behavioral consequences of the pandemic, rather than an exclusive outcome of the harvesting effect. In this respect, the recent re-explosion of cases and hospitalizations in the area of Brescia in the second half of February 2021, which led to the rapid imposition of *ad hoc* more severe restrictive measures, is a telltale sign that COVID-19 did not exhaust its impetus with the first wave in these territories.

Finally, excess mortality estimates computed over the entire February - November 2020 period confirm that subsequent reductions did not counterbalance the initial boom in Northern Italy mortality. Indeed, total excess mortality over this time-span is still 20% above what would have happened under 'ordinary' conditions.

Two caveats are in order regarding the credibility of our findings. While COVID-19 incidence was extremely low throughout the summer in Italy, including Northern regions, there is a possibility that many COVID-19 survivors from the first wave may have been fatally weakened by the virus and died several months later (Canoui-Poitrine et al., 2021). We acknowledge that this mechanism may be at play, but at the same time, we do not deem it to be so substantial to significantly alter the overall mortality trend, let alone reverse the sign of the excess mortality detected. Second, we only focus on the very short-run. Even though the harvesting effect is intrinsically a short-run phenomenon, the few months for which we have data may not be sufficient for the reabsorption to arise, and mortality displacement could take place over a longer time span.

Still, the very limited presence of COVID-19-induced mortality displacement in the short-run makes the health costs of the pandemic even more dramatic, suggests that COVID-19 can significantly shorten life expectancy, and restates once more the case for containment policies aimed at minimizing as much as possible the sanitary emergency and the death toll of the pandemic.

Our evidence is indeed preliminary. We look at a circumscribed area, Northern Italy, and focus on the very short-run due to current data availability. Further research should extend this type of analysis to other parts of Italy, other countries, and other waves of the current pandemic. When COVID-19 is eventually brought under control, it will be possible to provide a definitive answer on whether the pandemic played a significant anticipatory role and triggered a substantial mortality displacement or not. For the moment, the answer seems to be no.